\documentclass[aps,preprint]{revtex4}

\usepackage{epsfig}
 
\def\bq{\begin{equation}} 
\def\eq{\end{equation}} 
 
\newcount\minutes 
\newcount\scratch 
 
\def\timestamp{%
\scratch=\time 
\divide\scratch by 60 
\edef\hours{\the\scratch} 
\multiply\scratch by 60 
\minutes=\time 
\advance\minutes by -\scratch 
---$\,$\hours:\null 
\ifnum\minutes< 10 0\fi 
\the\minutes}

\begin{document}

\preprint{\bf MADPH-04-1370}
\preprint{March 2004}

\vspace*{.5in}

\title{QCD Corrections to Jet Correlations in Weak Boson Fusion}

\author{Terrance Figy and Dieter Zeppenfeld}
\affiliation{Department of Physics, University of Wisconsin, Madison, WI
\vspace*{.5in}}

\begin{abstract}
Higgs boson production via weak boson fusion is sensitive to 
the tensor structure of the $HVV$ ($V=W,Z$) couplings, which distinguishes
loop induced vertices from SM expectations.
At the CERN Large Hadron Collider this information shows up most clearly
in the azimuthal angle correlations of the two forward and backward quark 
jets which are typical for weak boson fusion. We calculate the next-to-leading
order QCD corrections to this process, in the presence of anomalous
$HVV$ couplings. Gluon emission does not significantly change the 
azimuthal jet correlations. 
\end{abstract} 


\maketitle

{\it Introduction.}
The production of Higgs bosons in the weak boson fusion (WBF) process will
provide a direct and highly sensitive probe of $HWW$ and $HZZ$ couplings 
at the CERN Large Hadron Collider 
(LHC)~\cite{wbfhtautau,wbfhtoww,wbfhtophoton,Zeppenfeld:2000td,Asai:2004ws}.
The determination both of the strength and of the tensor structure of these 
couplings is crucial for the identification of the produced boson as
a remnant of the spontaneous symmetry breaking process which is responsible 
for $W$ and $Z$ mass generation. 

Within spontaneously broken, renormalizable gauge theories like the 
standard model (SM), this coupling originates from the kinetic energy term, 
$(D_\mu\Phi)^\dagger (D^\mu\Phi)$, of a scalar Higgs field, $\Phi$, whose 
neutral component obtains a vacuum expectation value (vev), 
$\Phi^0\to (v+H)/\sqrt{2}$. This replacement then leads to a characteristic
coupling in the interaction Lagrangian, of the form $HV_\mu V^\mu$ 
($V=W,Z$). The existence of the vev is necessary to
produce a trilinear $HVV$ coupling at tree level: with $v=0$ all couplings
to the gauge fields $V$ contain two scalar fields, i.e., only $HHV$
and $HHVV$ couplings would be generated. A trilinear $HVV$ coupling may also
be loop-induced, however. The SM $H\gamma\gamma$ and $Hgg$ effective couplings
are an example: they are induced by $W$-boson and/or top quark loops. Gauge 
invariance dictates a different tensor structure of these loop-induced
couplings: the corresponding effective Lagrangian contains the 
square of the field strength, i.e. the lowest order loop-induced terms
are of the form $HV_{\mu\nu}V^{\mu\nu}$ or 
$HV_{\mu\nu}\tilde V^{\mu\nu}$, where $\tilde V^{\mu\nu} = \frac{1}{2}
\epsilon^{\mu\nu\rho\sigma}V_{\rho\sigma}$ denotes the dual field strength
of the gauge field. 

The task of future Higgs experiments is, then, twofold: (i) to measure
the overall strength of the $HVV$ coupling, and (ii) to identify its
tensor structure. One would expect a loop-induced coupling to be much smaller
than the expected SM $HVV$ coupling strength. However, the measurement of 
WBF rates alone will not be sufficient to establish $H$ as being related to
spontaneous symmetry breaking: to give just two examples, the loop-induced 
couplings might be substantially
enhanced by additional non-SM particles in the loop or by the existence
of multiplets of large weak isospin which couple strongly to $H$. Or a
particular LHC signature may be strongly enhanced by a much larger $H$
decay branching ratio than in the SM. A confirmation that
the $HVV$ coupling has tree level strength is, thus, ambiguous: a clear
identification of the Higgs boson also requires the identification of
the tensor structure of the $HVV$ vertex.

It was pointed out some time ago that the azimuthal angle correlations of the
two quark jets in the weak boson fusion process $qQ\to qQH$ provide 
tell-tale signatures for the tensor structure of the $HVV$ 
couplings~\cite{Plehn:2001nj}: the SM expectation is for a flat 
distribution, while the loop-induced couplings lead to a pronounced dip
at azimuthal separations $\phi_{jj}$ of the two tagging jets 
of 90 degrees for a $HV_{\mu\nu}V^{\mu\nu}$ coupling and at
0 and 180 degrees for the CP violating $HV_{\mu\nu}\tilde V^{\mu\nu}$ vertex.
Observation of the tagging jets is 
crucial for isolating the WBF process from backgrounds and, therefore, 
their distributions will be available for all WBF samples. 
Also, signal to background ratios for WBF processes are expected to be 
very good within the SM, exceeding the 1:1 level for wide ranges of the
Higgs boson 
mass~\cite{wbfhtautau,wbfhtoww,wbfhtophoton,Zeppenfeld:2000td,Asai:2004ws}.

The analysis of Ref.~\cite{Plehn:2001nj} was performed at leading order (LO)
in QCD. This means that additional gluon emission, which might lead to a
de-correlation of the tagging jets, was ignored in the analysis. Subsequently
it was argued~\cite{Odagiri:2002nd} that such de-correlation effects play an 
important role in a related process, $gg\to Hgg$, when the two tagging jets 
are widely separated in rapidity, which is a typical requirement for WBF 
studies. In this Letter we analyze this question, by calculating the tagging
jet distributions in  next-to-leading order (NLO) QCD, for the production 
of a scalar $H$ via WBF with an arbitrary tensor structure of the $HVV$ vertex. 
If de-correlation is important, it should show up in the form of large
radiative corrections at NLO. We use the term ``Higgs boson'' as a generic name 
for the produced scalar in the following.

{\it The NLO calculation.}  
Our calculation is an extension of the NLO QCD 
corrections for the SM WBF processes $qQ\to qQH$ 
(and crossing related ones)~\cite{WBF_NLO,Figy:2003nv,Berger:2004pc}.  
For the total cross section these corrections have 
been known for over a decade~\cite{WBF_NLO}. Recently, we have recalculated 
them by developing a NLO parton level Monte Carlo program~\cite{Figy:2003nv}
which provides the
flexibility to calculate arbitrary distributions at NLO, such as the 
azimuthal angle correlations that we are interested in here. 

\begin{figure}[t] 
\centerline{ 
\epsfig{figure=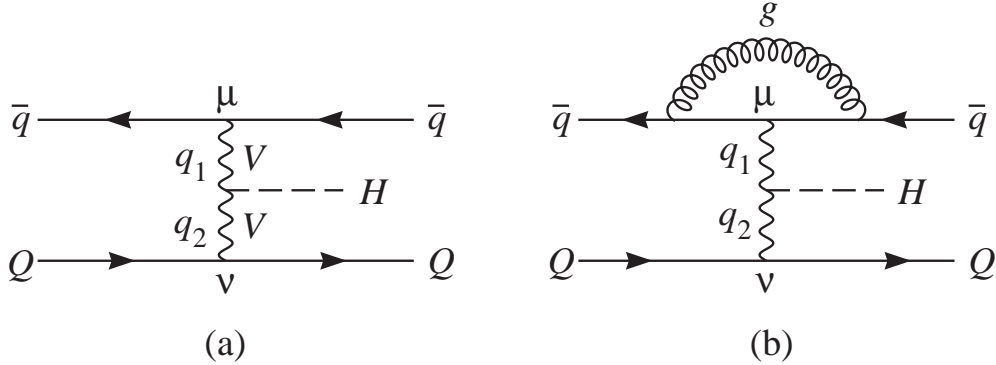,width=0.8\textwidth,clip=} \ \  
} 
\caption{ 
\label{fig:feyn2} 
Feynman graphs contributing to $\bar qQ\to \bar qQH$ at
(a) tree level and (b) including virtual corrections to the upper quark line.
The momentum labels and Lorentz indices for the internal weak bosons 
correspond to the vertex function of Eq.~(\ref{eq:vertex}).
}
\end{figure} 

The calculation of Ref.~\cite{Figy:2003nv} uses a SM vertex function,
$T^{\mu\nu}(q_1,q_2) = {2 m_V^2\over v} g^{\mu\nu}$ for the $HVV$ vertex in
Fig.~\ref{fig:feyn2}. Here we need to generalize this vertex to the most
general structure compatible with Lorentz invariance. Taking into account
that the quark currents in Fig.~\ref{fig:feyn2} and for the corresponding 
gluon emission processes are conserved, all terms proportional to 
$q_1^\mu$ or $q_2^\nu$ may be dropped, and the most general $HVV$ vertex 
may be written as
\begin{equation}
T^{\mu \nu}(q_{1},q_{2}) = a_1(q_1,q_2)\; g^{\mu \nu} + 
a_2(q_1,q_2)\; [q_{1}\cdot q_{2} g^{\mu \nu} - q_{2}^{\mu}q_{1}^{\nu}] + 
a_3(q_1,q_2)\; \varepsilon^{\mu \nu \rho \sigma} q_{1 \rho} q_{2 \sigma} \;.
\label{eq:vertex}
\end{equation} 
Here $q_1$ and $q_2$ are the four-momenta of the two weak bosons,
and the $a_i(q_1,q_2)$ are Lorentz-invariant form factors, which might,
for example, represent scalar loop integrals in a perturbative calculation.
It is straightforward to implement the general vertex of Eq.~(\ref{eq:vertex})
into our NLO QCD Monte Carlo: the virtual amplitude of Fig.~\ref{fig:feyn2}
is proportional to the Born amplitude, ${\cal M}_{\rm Born}$, irrespective 
of the structure of the $HVV$ vertex. Thus, all amplitudes reduce to a simple 
contraction of quark (or quark-gluon) currents with the vertex function
of Eq.~(\ref{eq:vertex}). 
These currents, and their contractions, are evaluated 
numerically, using the amplitude formalism of Ref.~\cite{HZ}. All other
aspects of the present NLO calculation are handled as in 
Ref.~\cite{Figy:2003nv}, except that we do not simulate any Higgs boson 
decays in the following. Factorization and renormalization scales are
fixed to $\mu_F=\mu_R=Q_i$ for QCD corrections to the first or second
quark line in Fig.~\ref{fig:feyn2}. Here $Q_1$ and $Q_2$ are the virtualities 
of the exchanged weak bosons. We use CTEQ6M parton 
distributions~\cite{cteq6} with $\alpha_s(M_Z)=0.118$ for all NLO results 
and CTEQ6L1 parton distributions for all leading
order cross sections. 

{\it Anomalous couplings and form factors.}
While the $g^{\mu \nu}$-term in the vertex function (\ref{eq:vertex})
corresponds to a SM Higgs coupling, the anomalous coupling terms 
$a_2$ and $a_3$ can be related to higher dimensional operators in an 
effective Lagrangian. They first appear at the dimension-5 
level~\footnote{The dimension 5 language is appropriate for, e.g., an isosinglet
scalar resonance $H$. For a Higgs doublet $\Phi$ with a vev, the leading operators
appear at dimension 6 level~\cite{Buchmuller:1985jz,Hagiwara:1993ck} and the 
couplings in Eq.~(\ref{eq:Leff}) are suppressed by an additional factor
$g_5^{HVV}\sim v/\Lambda$.}
and may be written as 
\begin{eqnarray}
\mathcal{L}_{5} &=& 
\frac{g^{HWW}_{5e}}{\Lambda_{5e}}  HW^{+}_{\mu \nu}W^{-\mu \nu} + 
\frac{g^{HWW}_{5o}}{\Lambda_{5o}}  H \tilde{W}^{+}_{\mu \nu}W^{-\mu \nu} + 
\nonumber \\  &&
\frac{g^{HZZ}_{5e}}{2\Lambda_{5e}} HZ_{\mu \nu}Z^{\mu \nu} + 
\frac{g^{HZZ}_{5o}}{2\Lambda_{5o}} H \tilde{Z}_{\mu \nu}Z^{\mu \nu}\;,
\label{eq:Leff}
\end{eqnarray}
where the subscript $e$ or $o$ refers to the CP even or odd nature of the 
individual operators.
In our discussion we will neglect possible contributions from 
$H \gamma \gamma$ and $H \gamma Z$ couplings which can appear in 
$SU(2)\times U(1)$ invariant
formulations~\cite{Buchmuller:1985jz,Hagiwara:1993ck}. 
The precise mix of $HWW$, $HZZ$, $HZ\gamma$ and $H\gamma\gamma$ contributions
is quite irrelevant for the observable azimuthal angle distributions, as 
long as we do not consider interference effects between SM and anomalous 
vertices, and it will not affect our conclusions about the size of
NLO corrections. For simplicity we therefore set $a_1=0$ for the 
anomalous coupling case and choose 
relative contributions from $WW$ and $ZZ$ fusion as in the SM, by 
taking $g^{HWW}_{5o} = g^{HWW}_{5e} = 1$,
$g^{HZZ}_{5e}=g^{HZZ}_{5o}=1/{\cos^{2}\theta_{W}}$, and by using either
$\Lambda_{5e} \simeq $480~GeV, $\Lambda_{5o}=\infty$ for the CP even case
or $\Lambda_{5o} \simeq $480~GeV, $\Lambda_{5e}=\infty$ for the CP odd case,
which roughly reproduces SM rates for a scalar mass of $m_H=120$~GeV.

The effective Lagrangian of Eq.~(\ref{eq:Leff}) produces couplings
\bq
a_2(q_1,q_2) = -\frac{2}{\Lambda_{5e}} g^{HWW}_{5e}\;, \qquad\qquad
a_3(q_1,q_2) = \frac{2}{\Lambda_{5o}} g^{HWW}_{5o} 
\eq
for the $HWW$ vertex, and
\bq
a_2(q_1,q_2) = -\frac{2}{\Lambda_{5e}} g^{HZZ}_{5e}\;, \qquad\qquad
a_3(q_1,q_2) = \frac{2}{\Lambda_{5o}} g^{HZZ}_{5o} 
\eq
for the $HZZ$ vertex. In general, the $a_i$ are form factors which are 
expected to be suppressed once the momentum transfer, $\sqrt{-q_i^2}$, 
carried by the 
virtual gauge boson reaches the typical mass scale, $M$, of the new physics
which is responsible for these anomalous couplings. Below we use the simple
ansatz
\begin{equation}
a_i(q_1,q_2) = a_i(0,0)\; \frac{M^2}{q_1^2-M^2}\;\frac{M^2}{q_2^2-M^2}
\label{eq:ff}
\end{equation}
for discussing the consequences of such form factor effects.

{\it Results:}
The typical signature of a weak boson fusion event at the LHC consists of the
two quark jets (tagging jets) and the Higgs decay products. The tagging jets
tend to be widely separated in rapidity, with one quite forward (typical
pseudorapidity of 3 to 4) and the second one backward, but frequently still
located in the central detector (pseudorapidity below 2.5). Various Higgs 
decay modes have been considered in the literature for WBF, 
$H\to WW$~\cite{wbfhtoww}, $H\to\tau\tau$~\cite{wbfhtautau}, and
$H\to\gamma\gamma$~\cite{wbfhtophoton} being the most promising ones. 
While optimized event selection varies, in particular for the decay products,
the cuts on the tagging jets are fairly similar in all analyses. 
Since here we are interested in the QCD features of WBF events, which do not 
depend on the Higgs decay mode, we perform our NLO analysis without simulating
Higgs decays, and we only impose typical WBF cuts on the tagging jets.

\begin{figure}[thb] 
\centerline{ 
\epsfig{figure=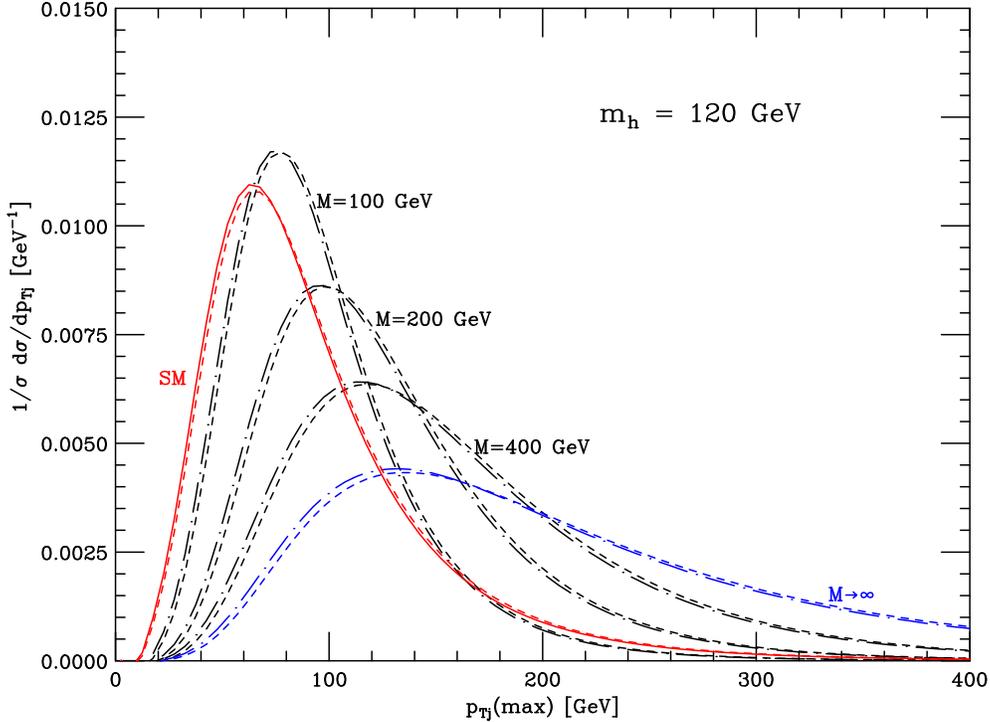,angle=90,width=0.8\textwidth} \ \  
} 
\caption{ 
\label{fig:ptj} 
Normalized transverse momentum distribution of the hardest jet for the 
SM Higgs boson (solid red line) and a scalar $H$ of mass $m_H=120$~GeV with
CP even anomalous coupling $a_2(q_1,q_2)$. The dash-dotted curves
correspond to different form factor scales 
$M=100$, 200, 400 GeV in Eq.~(\ref{eq:ff}) and $a_2=const.$ (blue curves)
at NLO. LO curves are shown by the dashed lines and differ very little 
from the NLO results.
}
\end{figure} 

In order to reconstruct jets from the final-state partons,
the $k_T$-algorithm~\cite{kToriginal} as described in Ref.~\cite{kTrunII} 
is used, with resolution parameter $D=0.8$. In a given event, the tagging 
jets are then defined as the two jets with the highest transverse momentum, 
$p_{Tj}$, with 
\bq
\label{eq:ptj_yj}
p_{Tj} \geq 20~{\rm GeV} \, , \qquad\qquad |y_j| \leq 4.5 \, .
\eq
Here $y_j$ denotes the rapidity of the (massive) jet momentum which is 
reconstructed as the four-vector sum of massless partons of 
pseudorapidity $|\eta|<5$. 
Backgrounds to weak-boson fusion are significantly suppressed by requiring
a large rapidity separation of the two tagging jets. This motivates the 
final cut 
\bq
\label{eq:rapgap}
\Delta y_{jj}=|y_{j_1}-y_{j_2}|>4\;, \qquad y_{j_1}\cdot y_{j_2} <0\;,
\eq
which includes the requirement that the two tagging jets reside in opposite 
detector hemispheres.

The structure of the $HVV$ coupling affects the production dynamics of $H$
and we can expect significant deviations in jet observables if, instead 
of the SM, anomalous couplings describe the vertex of Eq.~(\ref{eq:vertex}).
One example is shown in Fig.~\ref{fig:ptj}, where transverse momentum
distributions, $d\sigma/dp_{Tj}({\rm max})$, are compared between the SM 
(solid line) and  the CP even coupling $a_2(q_1,q_2)$, with different form 
factor scales $M$ in Eq.~(\ref{eq:ff}). Here, $p_{Tj}({\rm max})$ is the 
maximum $p_T$ of the two tagging jets. Only the shape of the distribution
is considered, since the rate can always be adjusted by multiplying the 
anomalous couplings by a constant factor. Also, we should note that a 
CP odd coupling leads to very similar curves for a given form factor scale.
In all cases we show the LO expectations (dashed lines) together with the 
NLO results: QCD corrections are of order 10\%, typically, and well under 
control.

One finds that anomalous $HVV$ couplings generally lead to harder $p_T$ 
spectra of the two tagging jets. Since the anomalous Lagrangian in
Eq.~(\ref{eq:Leff}) couples the Higgs boson to weak boson field strengths,
transverse polarizations of the incident $VV$ pairs dominate the anomalous
case, while longitudinal $VV$ fusion is responsible for SM Higgs production.
A telltale sign of transverse vector boson fusion is the more central and,
hence, higher $p_T$ production of the tagging jets. This effect is enhanced 
by the momentum factors in the $HVV$ anomalous vertices.

While the changed transverse momentum distributions in Fig.~\ref{fig:ptj}
could be used to rule out the SM, the reverse is not readily possible: a jet 
transverse momentum distribution compatible with SM expectations might be
faked by anomalous couplings and a judiciously chosen form factor behavior
of the coefficient functions $a_2$ or $a_3$ in Eq.~(\ref{eq:ff}). The 
different scale choices in Fig.~\ref{fig:ptj} demonstrate this effect: a low
form factor scale of $M=100$~GeV or slightly lower would be difficult to
distinguish from the SM expectation and one can certainly find a functional
form of the form factors which reproduces the SM within experimental errors.

\begin{figure}[thb] 
\centerline{ 
\epsfig{figure=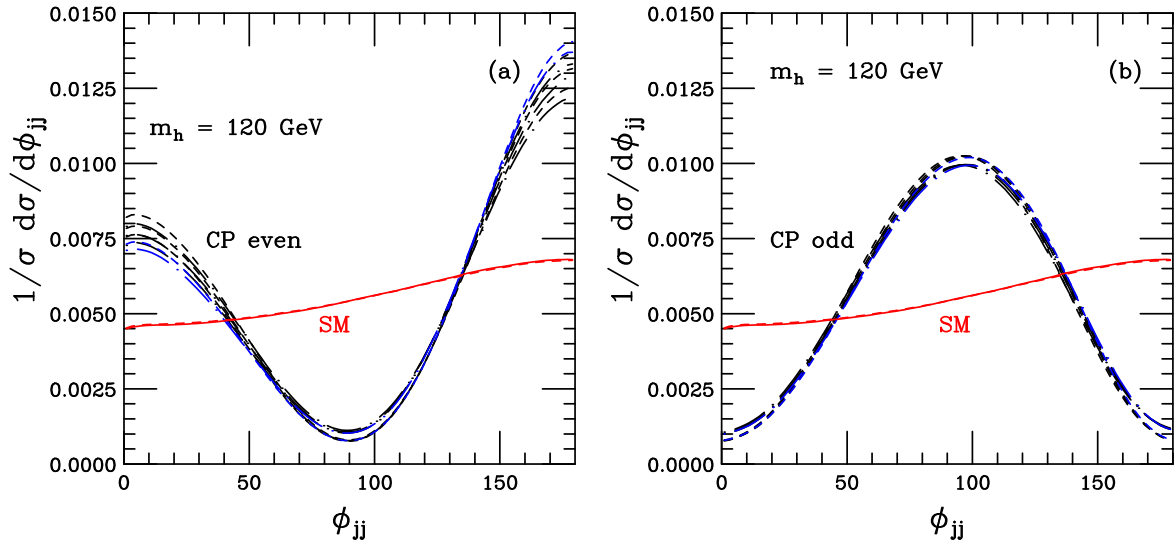,angle=90,width=0.95\textwidth,clip=} \ \  
} 
\caption{
\label{fig:phijj} 
Normalized azimuthal angle distribution, $1/\sigma\;d\sigma/d\phi_{jj}$
where $\phi_{jj}$ is the azimuthal angle separation of the two tagging jets.
NLO (solid and dot-dashed) and LO results (dashed lines) are shown for 
$m_H=120$~GeV in the SM (red curves) and (a) for a CP even anomalous coupling 
$a_2(q_1,q_2)$, (b) for a CP odd  anomalous coupling $a_3(q_1,q_2)$
with form factor scales $M=100$, 200, 400 GeV and (blue curves) $M=\infty$.
}
\end{figure} 

A much better observable for distinguishing the different 
tensor structures of the $HVV$ vertex is the azimuthal angle correlation of 
the two tagging jets, $d\sigma/d\phi_{jj}$~\cite{Plehn:2001nj}. 
Here $\phi_{jj}$ is the azimuthal angle between the two tagging jets. 
The corresponding distributions are shown in Fig.~\ref{fig:phijj} for the 
SM (solid line) and for the same choices of form factors as before.
The dip at $\phi_{jj}=90$ degrees for the CP even coupling 
and the suppression at 0 and 180 degrees for the CP odd coupling are clean
signatures which only depend on the tensor structure of the couplings and 
not on the precise dynamics which is responsible for the form factors.
The remaining form factor dependence is very small and can be explained by 
kinematic effects related to the higher average jet transverse momentum for big form 
factor scales, $M$: at small $\phi_{jj}$ two high $p_T$ jets recoil against the 
$H$ scalar, resulting in an increased invariant mass of the event compared 
to the situation with two back-to-back jets. This leads to a more asymmetric 
$\phi_{jj}$ distribution for high form factor scales.

The pronounced dip at 90 degrees, which is characteristic of the CP even 
coupling, is also found in $Hjj$ production via gluon fusion~\cite{ggh},
at LO. This is not surprising because, in the large top mass limit, the 
$Hgg$ vertex can be described by an effective Lagrangian proportional to
$HG_{\mu\nu}^aG^{a\mu\nu}$, which exhibits the same field strength squared 
behavior and hence the same tensor structure as the CP even $HVV$ coupling 
in Eqs.~(\ref{eq:vertex},\ref{eq:Leff}). Since the two tagging jets are 
far apart from each other, separated by a large rapidity gap of 4 units
of rapidity or more, this LO behavior may be significantly reduced by 
gluon radiation when higher order QCD corrections are taken into account.
Such de-correlation effects have been studied for dijet events at the 
Tevatron~\cite{tevatron}. For $Hjj$ production via gluon fusion,
Odagiri~\cite{Odagiri:2002nd} has argued that the dip structure is largely 
washed out by additional gluon emission between the two tagging jets.

\begin{figure}[htb] 
\centerline{ 
\epsfig{figure=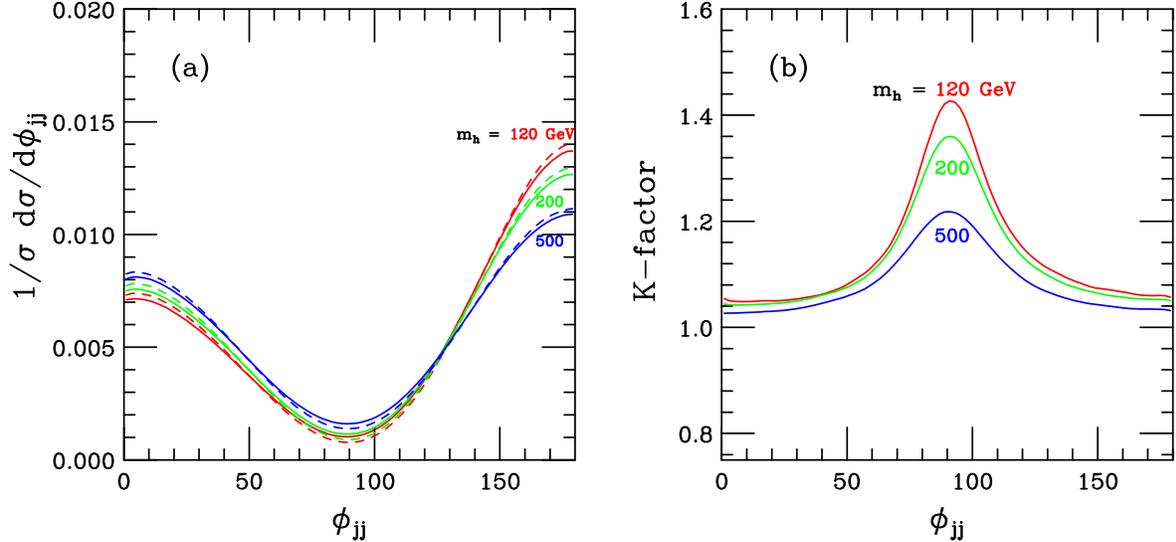,angle=90,width=0.95\textwidth,clip=} \ \  
} 
\caption{
\label{fig:phijjK}
Higgs mass dependence of the azimuthal angle separation $\phi_{jj}$ of 
the two tagging jets. In (a) the normalized azimuthal angle distributions
are shown at LO (dashed lines) and NLO (solid lines) for Higgs masses
of $m_H=$~120, 200, 500 GeV and a constant CP even anomalous coupling
$a_2$. Corresponding K-factors are shown in (b).
}
\end{figure} 

Our NLO calculations show that such de-correlation effects are irrelevant for
weak boson fusion, where $t$-channel color singlet exchange severely suppresses
gluon radiation in the central region. The LO and the NLO curves in 
Fig.~\ref{fig:phijj} are virtually indistinguishable. In order to better 
exhibit the size of NLO QCD effects for the WBF case, we show, in 
Fig.~\ref{fig:phijjK}(a) the azimuthal angle correlations for a pure CP even
anomalous coupling for three different Higgs masses, $m_H=120$, 200 and 
500~GeV. Only small changes are visible when going from LO (dashed lines)
to NLO (solid lines). The differences between LO and NLO are smaller than 
kinematical effects that can be induced by cuts on the Higgs decay products
or by variations of the Higgs boson mass.

The small to modest size of the QCD corrections is quantified in 
Fig.~\ref{fig:phijjK}(b) where the $K$ factor for the distribution is
shown, which is defined as
\bq
K(\phi_{jj}) = \frac{d\sigma^{NLO}/d\phi_{jj}}{d\sigma^{LO}/d\phi_{jj}} \;.
\eq
The $K$-factor is below $\approx 1.4$ even in the dip region,
where the cross section is severely suppressed. Virtually identical
results hold for the CP-odd case. Clearly, the characteristic azimuthal 
angle distributions of the jets in WBF are not affected in any significant 
way by NLO QCD corrections.

{\it Conclusions:}
We have performed a first calculation of the NLO QCD corrections to
Higgs boson production via WBF in the presence of arbitrary anomalous 
$HVV$ ($V=W,Z$) couplings. Anomalous couplings lead to characteristic
changes in the azimuthal angle correlation of the two tagging jets 
in weak boson fusion events at the LHC, which provides for a very 
sensitive test of the 
tensor structure of the $HVV$ couplings of the Higgs boson
or of any other scalar with sufficiently large production cross section in
WBF~\cite{Plehn:2001nj}. 
We have shown by explicit calculation that these azimuthal correlations
are not washed out by gluon emission, at NLO QCD, even though the tagging 
jets are widely separated in rapidity. 
This behavior can be 
understood as a consequence of $t$-channel color singlet exchange in WBF
which severely suppresses the central gluon radiation which 
might cause tagging jet de-correlation.

\section*{Acknowledgments}
This research was supported in part by the University
of Wisconsin Research Committee with funds granted by the Wisconsin Alumni
Research Foundation and in part by the U.S.~Department of Energy under
Contract No.~DE-FG02-95ER40896.

\end{document}